\documentstyle[pra,aps,preprint]{revtex}
 \begin{document}
 \draft
 \title{Some Cosmological Applications of Two Measures Theory}

\author{E. I. Guendelman \thanks{guendel@bgumail.bgu.ac.il} and
A. B. Kaganovich \thanks{alexk@bgumail.bgu.ac.il}}
\address{Physics Department, Ben Gurion University of the Negev, Beer
Sheva
84105, Israel}
\maketitle
\begin{abstract}
Scale invariance is considered  in          
the context of a gravitational theory where                                   
the action, in the first order formalism, is of the form $S =                 
\int L_{1} \Phi d^4x$ + $\int L_{2}\sqrt{-g}d^4x$ where $\Phi$ is a           
density built out of degrees of freedom independent of the metric.            
For global scale invariance, a "dilaton"                                      
$\phi$ has to be introduced, with non-trivial potentials $V(\phi)$ =          
$f_{1}e^{\alpha\phi}$ in $L_1$ and $U(\phi)$ = $f_{2}e^{2\alpha\phi}$ in      
$L_2$. In the effective Einstein frame, this leads to a non-trivial  
$\phi$ potential (of the Morse type)           
 which has a flat region with energy density $f_{1}^{2}/4f_{2}$
as $\phi\rightarrow\infty$. The addition of an $R^{2}$ term produces 
an effective potential with two connected flat regions: one of the 
Planck scale, that can be responsible for early inflation, and another 
for the description of the present universe.  
\end{abstract}

\bigskip
Here we study a generally coordinate invariant theory where 
in addition to the action term with the                                                   
ordinary measure of integration $\sqrt{-g}d^{4}x$, the action contains
  another one with the measure              
$\Phi d^{4}x$, where $\Phi$ is a density built out of degrees of freedom      
independent of the metric (we call this a "Two Measures Theory" (TMT)).                                                    
        For example, given 4-scalars $\varphi_{a}$ ($a =                       
1,2,3,4$), one can construct the density                                       
$\Phi =  \varepsilon^{\mu\nu\alpha\beta}  \varepsilon_{abcd}                   
\partial_{\mu} \varphi_{a} \partial_{\nu} \varphi_{b} \partial_{\alpha}       
\varphi_{c} \partial_{\beta} \varphi_{d}$.                                      
        One can allow both geometrical                                        
objects to enter the theory and consider\cite{GK1}                                  
\begin{equation}                                                              
S = \int L_{1} \Phi  d^{4} x  +  \int L_{2} \sqrt{-g}d^{4}x                   
\end{equation}                                                                
         Here $L_{1}$ and $L_{2}$ are                                         
$\varphi_{a}$  independent. There is a good reason not to consider            
mixing of  $\Phi$ and                                                         
$\sqrt{-g}$ , like                                                            
for example using                                                             
$\frac{\Phi^{2}}{\sqrt{-g}}$. This is because (1) is 
invariant (up to the integral of
a total
divergence) under the infinite dimensional symmetry                           
$\varphi_{a} \rightarrow \varphi_{a}  +  f_{a} (L_{1})$                       
where $f_{a} (L_{1})$ is an arbitrary function of $L_{1}$ if $L_{1}$ and      
$L_{2}$ are $\varphi_{a}$                                                     
independent. Such symmetry (up to the integral of a total divergence) is      
absent if mixed terms are present.

        We will study now the dynamics of a scalar field $\phi$ interacting   
with gravity as given by the action (1) with\cite{G1}                          
\begin{equation}                                                              
L_{1} = -\frac{1}{\kappa} R(\Gamma, g) + \frac{1}{2} g^{\mu\nu}               
\partial_{\mu} \phi \partial_{\nu} \phi - V(\phi), \quad  L_{2} = U(\phi)           
\end{equation}                                                                
                                                                              
\begin{equation}                                                              
R(\Gamma,g) =  g^{\mu\nu}  R_{\mu\nu} (\Gamma) , \  R_{\mu\nu}                   
(\Gamma) = R^{\lambda}_{\mu\nu\lambda}, \ 
 R^{\lambda}_{\mu\nu\sigma} (\Gamma)   
= \Gamma^{\lambda}_                                                           
{\mu\nu,\sigma} +                          
\Gamma^{\lambda}_{\alpha\sigma}  \Gamma^{\alpha}_{\mu\nu} -                   
(\nu\leftrightarrow\sigma)
\end{equation}

        In the variational principle, $\Gamma^{\lambda}_{\mu\nu}$,
$g_{\mu\nu}$, the measure scalar fields                                          
$\varphi_{a}$ and the  scalar field $\phi$ are all to be treated              
as independent variables.                                                     

The action (1),(2) is invariant under 
the global scale transformation ($\theta$ =             
constant)                                                                     
\begin{equation}                                                              
g_{\mu\nu}\rightarrow e^{\theta}g_{\mu\nu}, \                             
\phi \rightarrow\phi - \frac{\theta}{\alpha}, \
\varphi_{a}\rightarrow\lambda_{a} \varphi_{a} \ (no \, sum \, on\, a),\ 
\Phi \rightarrow \left(\prod_{a} {\lambda}_{a}\right)\Phi\equiv\lambda
\Phi , 
\end{equation}                                                                
where $\lambda = e^{\theta}$, provided  $V(\phi)$     
and $U(\phi)$ are of the form                                                                          
$V(\phi) = f_{1}  e^{\alpha\phi}$, \   $U(\phi) =  f_{2}                            
e^{2\alpha\phi}$.                                                               
In this case we call the     
scalar field $\phi$ needed to implement scale invariance "dilaton".           
                                                                           
      Let us consider the equations which are obtained from                 
the variation of the $\varphi_{a}$                                            
fields. We obtain then  $A^{\mu}_{a} \partial_{\mu} L_{1} = 0$                
where  $A^{\mu}_{a} = \varepsilon^{\mu\nu\alpha\beta}                         
\varepsilon_{abcd} \partial_{\nu} \varphi_{b} \partial_{\alpha}               
\varphi_{c} \partial_{\beta} \varphi_{d}$. If $\Phi\neq 0$ then                              
det $(A^{\mu}_{a}) =\frac{4^{-4}}{4!} \Phi^{3} \neq 0$        
and we obtain that $\partial_{\mu} L_{1} = 0$,          
 or that                                                                      
$L_{1}  = M$,                                                                 
where $M$ is a constant. The appearance of the integration constant $M$ 
spontaneously breaks the global scale symmetry. The constant $M$
enters also in a self-consistency            
condition of the equations of motion                                          
that allows us to solve for $ \chi \equiv \frac{\Phi}{\sqrt{-g}}
 = \frac{2U(\phi)}{M+V(\phi)}$.

        To get the physical content of the theory, it is convenient to go     
to the Einstein conformal frame where                                         
$\overline{g}_{\mu\nu} = \chi g_{\mu\nu}$.                                       
 In terms of $\overline{g}_{\mu\nu}$   the non       
Riemannian contribution 
disappears from all the equations, which can be written then in the Einstein      
form ($R_{\mu\nu} (\overline{g}_{\alpha\beta})$ =  usual Ricci tensor)        
\begin{equation}                                                              
R_{\mu\nu} (\overline{g}_{\alpha\beta}) - \frac{1}{2}                         
\overline{g}_{\mu\nu}                                                         
R(\overline{g}_{\alpha\beta}) = \frac{\kappa}{2} 
\left[\phi_{,\mu} \phi_{,\nu} - \frac{1}{2} \overline
{g}_{\mu\nu} \phi_{,\alpha} \phi_{,\beta} \overline{g}^{\alpha\beta}
+ \overline{g}_{\mu\nu} V_{eff} (\phi)\right]
\end{equation}                                                                
where                                                                         
$V_{eff} (\phi) = \frac{1}{4U(\phi)}  (V+M)^{2}$.                               
        If $V(\phi) = f_{1} e^{\alpha\phi}$  and  $U(\phi) = 
f_{2}e^{2\alpha\phi} $  
as required by scale invariance, we obtain 
\begin{equation}                                                              
        V_{eff}  = \frac{1}{4f_{2}}  (f_{1}  +  M e^{-\alpha\phi})^{2}        
\end{equation}

{ \it A minimum of $V_{eff}$ is achieved at zero         
cosmological constant} for the case $\frac{f_{1}}{M} < 0 $ at the point         
$\phi_{min}  = - \frac{1}{\alpha} \ln |\frac{f_1}{M}| $. Finally,        
the second derivative of the potential  $V_{eff}$  at the minimum is          
$V^{\prime\prime}_{eff} = \frac{\alpha^2}{2f_2} |f_{1}|^{2} > 0$        
if $f_{2} > 0$, so that a realistic scalar field potential, with           
massive excitations when considering the true vacuum state, is achieved in    
a way consistent with the idea of scale invariance.

The above simple model  contains
 a single flat region which can do a
good job at describing either early inflation or a slowly accelerated universe.
But what about explaining both, including a possible transition between
these two epochs?
A simple generalization of the action $S$ given by (2) will do this job.
This is the addition of a scale invariant term\cite{rsqr}
\begin{equation}
S_{R^{2}} = \epsilon  \int (g^{\mu\nu} R_{\mu\nu} (\Gamma))^{2} \sqrt{-g}
d^{4}x
\end{equation}
to the action given by Eqs.(1),(2).
In the first order formalism $ S_{R^{2}}$ is not
only globally scale invariant
but also locally scale invariant, that is conformally invariant (recall
that
in the first order formalism the connection is an independent degree of
freedom and it does not transform under a conformal transformation of
the metric).

If we follow the previous steps (when $R^{2}$ was absent) and study the model
in slow rolling approximation, we get the following effective potential
\begin{equation}
 V_{eff}  =
\frac{(f_{1} e^{ \alpha \phi }  +  M )^{2}}
{4(\epsilon \kappa ^{2}(f_{1}e^{\alpha \phi}  +  M )^{2} +
f_{2}e^{2 \alpha \phi })}
\end{equation}

The limiting values of $ V_{eff} $ are:
first, for asymptotically
large positive values, i.e. as $ \alpha\phi \rightarrow  \infty $,
we have
$V_{eff} \rightarrow
\frac{f_{1}^{2}}{4(\epsilon \kappa ^{2} f_{1}^{2} + f_{2})} $ ;
second, for asymptotically large but negative values of the scalar field
(i.e. as $\alpha \phi \rightarrow - \infty  $)  we have
$ V_{eff} \rightarrow \frac{1}{4\epsilon \kappa ^{2}}$ ,
which, if $\epsilon $
is a number of order one, means that we have an energy density determined
by the Planck scale. This region is suitable for supporting the early
inflation.
In spite of the unusual features of TMT, the $R^{2}$ term leads 
qualitatively to the same effect that was found by Starobinsky
in the quadratic in curvature theory he studied\cite{St}

Another possible generalization of the model consists of taking the most 
general form for the Lagrangians $L_{1}$ and $L_{2}$. For example one can
choose $L_{2}$ in the form similar to $L_{1}$ but with different 
coefficients in front of different terms. The corresponding models
(linear in the scalar curvature $R$) were studied in Refs.\cite{GK2,GK3} 
and they provide solution of a number of
puzzles,
like the origin of fermion generations, fifth force problem in
quintessence scenario and a new model of dark matter. 
The main results are reported in our other
contribution
to this section.


\begin{thebibliography}{99}

\bibitem{GK1} For a review see,
 E.I.Guendelman and A.B.Kaganovich, 
{\it Phys. Rev.} {\bf D60}, 065004 (1999).                     

\bibitem{G1} 
E.I. Guendelman, {\it Mod. Phys. Lett.} {\bf A14}, 1043 (1999);
ibid. {\bf A14}, 1397 (1999); 
gr-qc/9901067.                                   

\bibitem{rsqr}
E.I. Guendelman and O. Katz,
{\it Class. Quant. Grav} {\bf 20}, 1715 (2003). 

\bibitem{St}
A.A. Starobinsky, {\it Phys. Lett.} {\bf B91}, 99 (1980).

\bibitem{GK2}
E.I. Guendelman and A.B. Kaganovich, {\it Int. J. Mod. Phys.}
{\bf A17}, 417 (2002);  {\it Mod. Phys. Lett.} {\bf A17}.

\bibitem{GK3}
E.I. Guendelman and A.B. Kaganovich, gr-qc/0312006.


\end{thebibliography}
\end{document}